\documentclass{ws-ijmpa}

\def\1{{\bf 1}}
\def\[{\left[}
\def\]{\right]}
\def\be{\begin{eqnarray}}
\def\ee{\end{eqnarray}}
\def\nn{\nonumber}
\def\({\left(}
\def\){\right)}
\def\bk#1{\langle#1\rangle}
\def\eq#1{(\ref{#1})}
\def\o{\omega}

\def\C{{\cal C}}
\def\G{{\cal G}}

\def\L{{\cal L}}

\def\s{\sigma}

\def\edoc{\end{document}}
\def\M{\overline{M}}
\def\bp{\begin{pmatrix}}
\def\ep{\end{pmatrix}}
\def\1{{\bf 1}}
\begin{document}

%
%

\title{Horizontal Symmetry from the Bottom Up}

\author{C.S. Lam}

\address{Department of Physics, McGill University\\
Montreal, QC, Canada, and\\
Department of Physics and Astronomy, University of British Columbia\\
Vancouver, BC, Canada\\
Lam@physics.mcgill.ca}

%

\maketitle


\begin{abstract}
A general method to derive  horizontal symmetry from a mixing matrix is reviewed. The technique has been applied to deduce leptonic symmetry 
from the tri-bimaximal neutrino mixing matrix and three of its variations. The question of how the quark mixing can  be accommodated within the leptonic
symmetry group is discussed, including in this connection an example based on the group $D_4$.
\end{abstract}

\ccode{PACS numbers: 12.15.Ff}

\vspace{.5cm}
This talk consists of two parts: a review of my previously work\cite{1}, in which reference to other related research can be found, and some recent thoughts
of using the symmetry group $D_4$  to accommodate both leptonic and  quark mixings.

\section{Introduction}	
The existence of three almost identical generations of quarks and leptons  suggests  a new quantum number
to distinguish them, a quantum number that arises from a new symmetry commonly known as horizontal symmetry. In order to account for mass differences 
of the three generations and their mixing, the symmetry must be broken, usually assumed spontaneously. 
The standard approach is to pick a horizontal symmetry group $\G$, often
a finite group to avoid the appearance of Goldstone particles, assign irreducible representations under that group to various
left-handed and right-handed fermions, as well as the scalar particles (Higgs) needed to break the symmetry. Yukawa couplings invariant under $\G$
are constructed, vacuum expectation values of the Higgs are assigned to break the symmetry, and mass matrices are read out from the broken Lagrangian.
The mass matrices so obtained depend on various parameters: the Yukawa coupling constants and the vacuum expectation values. 
These parameters are then adjusted to fit the masses and experimental mixing parameters obtained from these mass matrices.

The regularity of neutrino mixing has aroused intense interest in recent years; various symmetry groups have been invoked to explain it. They  include $S_3, A_4, T', S_4, 
PSL(2,7), \Delta(2n^2), \Delta(6n^2), D_n, Q_n$, and others. Because of the presence of numerous adjustable parameters in each case, it is difficult to tell
how much of the success in explaining the mixing is due to the symmetry invoked, and how much is due to parameter fitting. In an attempt to settle this question,
I proposed some time ago two criteria, demanding that

\be \bf{ 
symmetry\ of\ the\ mass\ matrices\ should\ be}\qquad\nn \\ \bf{symmetries\ of\ the\ horizontal\ symmetry\ group\ \G,}\label{c1}\ee
and
\be \bf{
vacuum\ alignments\ should\ be\ assigned\ to\ satisfy\ \eq{c1}.}\label{c2}\ee

We will see
in the next section that the symmetry of the mass matrices (`residual symmetry') can be determined from the mixing matrix. Consequently
the minimal horizontal group $\G$ is, according to \eq{c1}, 
just the group generated by the residual symmetries obtained from the mixing matrix. Moreover, if symmetry breaking is carried out according
to \eq{c2}, then the mixing matrix is automatically recovered provided the residual symmetry is `non-degenerate'.
In that case, the remaining free parameters are the Yukawa coupling constants, and they are used exclusively to fit the masses. If the residual symmetry
is not `non-degenerate', then some free parameters may have to be used to fit the mixing matrix as well.

Given a mixing matrix, it should be noted that the horizontal symmetry $\G$ is not unique for three reasons.
First, if $\G$ satisfies \eq{c1}, then any group containing $\G$
also satisfies \eq{c1}. This kind of non-uniqueness is not serious because, for the sake of economy, one usually uses the smallest group anyway unless there is a good 
reason not to. Secondly, for reasons to be explained in the next section, the residual symmetry itself is not unique, but as long as it is `non-degenerate', we
will recover the mixing matrix automatically whatever the corresponding $\G$ is.  
Thirdly, the symmetry group $\G$ so deduced
is strictly speaking only the symmetry group of the left-handed fermions. It can however be taken to be the symmetry group of all the fermions and the
local Lagrangian, if we assign the right-handed fermions to transform like the left-handed fermions.

\section{Residual and Horizontal Symmetries}
Let $e,\nu,u,d$ denote the charged leptons, the neutrinos, the up-type and the down-type quarks, and $M_\alpha\ (\alpha=e,\nu,u,d)$ their corresponding $3\times 3$
Dirac mass matrices, connecting left-handed to right-handed fermions. It is important to note that fermion masses and mixings  can already be obtained  by diagonalizing the
left-handed to left-handed mass matrices $\M_a=M_aM_a^\dagger$ ($a=e,u,d$) and $\M_\nu=M_\nu M_N^{-1}M_\nu^T$,
assuming the active neutrinos to be given by a type-I see-saw mechanism, with a
right-handed Majorana mass matrix $M_N$. For that reason we cannot possibly deduce any information on the symmetry of
 the right-handed fermions just from masses and mixing. From now on we shall deal with the mass matrices $\M$ exclusively until the last section.

I shall concentrate on explaining how to obtain the residual symmetries of the left-handed leptons from the mixing matrix. The procedure in the
quark sector is similar. Let $U=(u_1,u_2,u_3)$ be the PMNS mixing matrix, with $u_i$ standing for  the $i$th column of $U$. It is the matrix to diagonalize
$\M_\nu$ in the basis where $\M_e$ is diagonal. Each $U$ gives rise to three different unitary symmetry operators $G_i$ of the neutrino mass matrix, satisfying
$G_i^T\M_\nu G_i=\M_\nu$, and given by the formula
\be
G_i=u_iu_i^T-(u_ju_j^T+u_ku_k^T)\quad (i=1,2,3),\label{G}\ee
where $j,k$ are the other two indices different from $i$.
For the mixing matrix in the tri-bimaximal form
\be
U={1\over\sqrt{6}}\begin{pmatrix} 2&\sqrt{2}&0\\ -1&\sqrt{2}&\sqrt{3}\\ -1&\sqrt{2}&-\sqrt{3} \end{pmatrix},\label{tribi}\ee
these three symmetry operators are
\be
G_1&=&{1\over 3}\begin{pmatrix}1&-2&-2\\ -2&-2&1\\ -2&1&-2\end{pmatrix},\quad
G_2=-{1\over 3}\bp 1&-2&-2\cr -2&1&-2\cr -2&-2&1\ep,\quad
G_3=-\bp 1&0&0\cr 0&0&1\cr 0&1&0\cr\ep.\label{G}\ee
A variance and relaxation of the tri-bimaximal mixing is to assume only one of its columns to be firmly known, with the other two parameterized subject only to unitarity.
If only the $i$th column of the tri-bimaximal matrix is  firmly known, then the residual symmetry is given by a single $G_i$. For $i=3$, bimaximal mixing
is present but trimaximal mixing may not. For $i=2$, trimaximal mixing is present but bimaximal mixing may not. For $i=1$, there is no good name for it but for
easiness of referral we will just call it `unimaximal mixing' for now.

We turn to the charged-lepton sector where $\M_e$ is diagonal, and denote its unitary symmetry opeators by $F$. They satisfy $F^\dagger \M_eF=\M_e$ and $F^\dagger F=1$. 
Any diagonal unitary matrix is a solution and hence a qualified symmetry operator. Let me divide these solutions into two categories, 
{\tt non-degenerate}, and {\tt degenerate}.
The former refers to  cases when all three diagonal entries of $F$ are different, and the latter refers to the situation in which two of the entries are the same.
The significance of this division will become clear in a moment.

Using \eq{c1}, a minimal $\G$ is simply the group generated by $F$ and $G$, where $F$ is one of the charged lepton residual symmetries, and $G$ denotes 
collectively all the known
residual symmetries in the neutrino sector. We can always obtain a larger $\G$ by taking several $F$'s as generators, but usually it is the smallerst that
we want.

If we take $F=F_3:={\rm diag}(1,\o,\o^2)$, with $\o=e^{2\pi i/3}$, then $F^3=1$, and the minimal horizontal symmetry groups can be shown to be
\begin{enumerate}
\item $\G=\{F_3,G_1,G_2,G_3\}=S_4$ for tri-bimaximal mixing.
\item $\G=\{F_3, G_1\}=S_4$ for unimaximal mixing.
\item $\G=\{F_3, G_2\}=A_4$ for trimaximal mixing.
\item $\G=\{F_3, G_3\}=S_3$ for bimaximal mixing.
\end{enumerate}
Moreover, it can be shown that as long as $F$ is non-degenerate, then $\G=\{F,G_1,G_2,G_3\}\supset S_4$. Let me stress that this theorem is based only on the 
physically reasonable criterion of \eq{c1}, and not some other criterion that has been incorrectly quoted in the literature.

\section{Effective Lagrangian}
Let $\L$ be the Lagrangian density invariant under some horizontal symmetry group. After integrating out the right-handed fermions, we obtain an effective
Lagrangian density $\L_{eff}$ consisting of the left-handed leptons $e_L, \nu_L$, the (possibly composite) Higgs fields $\Phi$ in the charged lepton section, 
and the Higgs fields $\Psi$ in the neutrino sector. $\L_{eff}(e_L,\nu_L,\Phi,\Psi)$ is invariant under the left-handed symmetry group $\G$
determine previously. Note that the effective Lagrangian may
not even be local, but that does not matter because all that concerns us is its invariance under the left-handed symmetry group $\G$.

\section{Recovering the Mixing Matrix}
\eq{c2} is used to determine the vacuum alignments when $\G=\{F,G\}$ is broken. If $\L_{eff}$ is to remain invariant under the residual symmetries, 
then to satisfy \eq{c2} the vacuum alignments
must obey
\be
F^A\bk{\Phi^A}=\bk{\Phi^A},\quad G^A\bk{\Psi^A}=\bk{\Psi^A}\label{fgvev}\ee
for every irreducible representation $A$ of $\G$, where $F^A, G^A$ are the irreducible representation of $F, G$, and $\Phi^A, \Psi^A$ are the irreducible representations
of $\Phi, \Psi$.

With the vacuum alignments satisfying \eq{fgvev}, the residual symmetries of the original mixing matrix are preserved. If $F$ is non-degenerate
and diagonal, then the original mixing
matrix is automatically recovered for the following reason. Since $\M_e$ commutes with $F$, the non-degeneracy of $F$ guarantees that $\M_e$ is diagonal when $F$ is.
$U$ is then recovered by filling its columns by the eigenvectors of $G$ with $+1$ eigenvalues.
Note that this is not necessarily correct if $F$ is degenerate, for then even in the representation when $F$ is diagonal, $\M_e$ may not be, and a further rotation
may be necessary to make it diagonal. In that case the $U$ so calculated is not yet the mixing matrix. For easiness of reference, we repeat this conclusion below:
\be \bf{ 
the\ mixing\ matrix\ is\ automatically\ recovered\ if\ F\ is\ non{\rm -}degenerate.}\label{c3}\nn\\ \label{c3}\ee

\section{Using $D_4$ to Accommodate Quark Mixing}
Quark mixing is irregular, has small mixing angles, is hence completely unlike leptonic mixing. Its left-handed
symmetry group obtained from \eq{c1} must also be very different from the lepton's. This is unpleasant, so it is natural to ask whether there is any way to
 accommodate quark mixing within the leptonic symmetry group $\G$. 

If one is willing to abandon \eq{c1} and/or \eq{c2} in the quark sector, that could presumably be done by 
fitting. One then chooses quark representations, new Higgs and new alignments if necessary, 
and use the tunable parameters to fit the quark data alone, or the quark and leptonic data together.
There are some fairly successful models in the literature of this kind, 
giving a decent fit to existing quark and leptonic data.
Since the fit is usually not perfect, it is hard to pinpoint the origin of the success, and in exactly what way the chosen group $\G$ 
and not some others is the real horizontal symmetry.

Alternatively,
one can try to
retain \eq{c1}, \eq{c2}, and the same set of Higgs fields, 
at the expense of giving up a full explanation of  quark mixing in the lowest order. 
For example, one might try to accommodate Cabibbo mixing in the lowest order, 
but leave out the smaller mixings with the third generation and attribute them to the effect of a higher order correction of vacuum alignments.  
In the rest of this section, we will discuss a simple example of this kind.

The first question to settle is, if we adopt the same group $\G$, use the same Higgs fields $\Phi, \Psi$ with the same vacuum alignments in the leptonic and  the quark sectors,
how come we do not up with the same mixing matrix for both? After all, according to \eq{c2} and \eq{fgvev}, we expect the same residual symmetry
to emerge from the same vacuum alignments, and according to \eq{c3}, same residual symmetry leads to the same mixing matrix.
The answer is that, whereas $\Phi$ is the Higgs for $e$ and $\Psi$ is the
Higgs for $\nu$ in the leptonic sector, the Higgs for $u$ and $d$ will both be $\Phi$ in the quark sector. Recall that the left-handed to left-handed mass
matrices $\M_\alpha$ is related to the Dirac mass matrices $M_\alpha$ by $\M_a=M_aM_a^\dagger$ for $a=e,u,d$, but because of the see-saw mechanism
for neutrino, the relation there is  $\M_\nu=M_\nu M_N^{-1}M_\nu^T$. It is therefore reasonable to choose the same Higgs $\bk{\Phi}$ for $e,u,d$, but a different one 
$\bk{\Psi}$ for $\nu$.
This however brings up another problem: 
if $\bk{\Phi}$ is used  for $u$ and $d$ both, we should end up with no quark mixing whatsoever if \eq{c3} applies. For that reason we will have to 
sidestep \eq{c3} by using a degenerate $F$
to allow Cabibbo mixing to be accommodated.

The simplest degenerate $F$ is given by one of the following three: 
\be F_a={\rm diag}(1,-1,-1),\quad F_b={\rm diag}(-1,1,-1),\quad F_c={\rm diag}( -1, -1, 1).\label{F}\ee
The left-handed leptonic symmetry groups generated by $F_a, F_b$, and/or $F_c$. and $G_1, G_2$, and/or $G_3$, can be computed in a straight forward manner. It turns out that 
if $G_1$ or $G_2$ is involved, the resulting group is infinite  no matter which  $F_{a,b,c}$ we choose. This leaves $G_3$ as the only possible residual symmetry
in the leptonic sector. Computation shows that $\{F_a, G_3\}=Z_2\times Z_2$, and both $\{F_b, G_3\}$ and $\{F_c, G_3\}$
give rise to the same group $D_4$,  the dihedral group with 8 elements, which contains both $F_b$ and $F_c$. 
In order to have a non-trivial mixing, we choose the non-abelian group $D_4$ as $\G$.

Before proceeding with the physics let us first summarize the relevant mathematical properties of the group. $D_4$ has five classes $\C_i$,
containing the following elements: $\C_1=(\1),\ \C_2=(F_{b}, F_{c}),\ \C_3=(G_3, G_3F_{a}),\ \C_4=(F_{c}G_3, G_3F_{c}),\ \C_5=(F_{a})$.
The group has four 1-dimensional irreducible representations, which we shall designate as $A, B, C, D$, and one 2-dimensional irreducible representation $E$. The
irreducible representations in the basis where $F_{b,c}$ are diagonal are given by the left-hand table of \eq{IR} below, in which $\s_i$ are the Pauli matrices. 
The defining representation of $G_3$ and $F_{a,b,c}$ shown in \eq{G} and \eq{F} can easily be seen to belong to $D\oplus E$. Since the residual symmetries
come from the mixing matrix of the left-handed leptons, the left-handed leptons themselves must also belong
to  $D\oplus E$, with the first generation in $D$, and the second and third generations together in $E$.
\be\begin{array}{|c|cccc|c|c|ccccc|} \cline{1-5} \cline{7-12} &F_{a}&F_{b}&F_{c}&G_3&\hspace{2cm}&&A&B&C&D&E\\ \cline{1-5} \cline{7-12}
A&1&1&1&1&\hspace{2cm}&A&A&B&C&D&E\\
B&1&-1&-1&1&&B&B&A&D&C&E\\
C&1&1&1&-1&&C&C&D&A&B&E\\
D&1&-1&-1&-1&&D&D&C&B&A&E\\
E&-\1&\s_3&-\s_3&-\s_1&&E&E&E&E&E&A+B+C+D\\
\cline{1-5} \cline{7-12} \end{array}\quad ,\label{IR}\ee
Both the character table and the Clebsch-Gordon (CG) series can be obtained from \eq{IR}. The allowed Clebsch-Gordan series are shown on the right-hand table of \eq{IR}.
If the two doublets in  $E\otimes E$ are $(a_1,a_2)$ and $(b_1,b_2)$, then 
\be A=a_1b_1+a_2b_2,\ B=a_1b_2+a_2b_1,\ C=a_1b_1-a_2b_2,\ D=a_1b_2-a_2b_1.\label{CGC}\ee

Now we have to decide whether $F_b$, or $F_c$, or both, is the residual symmetry in the $e$ sector. In order to be able to accommodate Cabibbo
mixing, I claim that has to be $F_c$ for the following reason. Once we pick a $F$, the alignment of $\bk{\Phi}$ is fixed by \eq{c2} and \eq{fgvev}.
Since we have decided to use the same $\bk{\Phi}$ in the $u$ and $d$ sectors as well, it compels us also to use this same $F$ to be the residual symmetry in both quark sectors. 
If the first and the second diagonal entries of $F$ are the same, which is the case for $F_c$, then \eq{c3} is no longer valid, and $M_a\ (a=u,d,e)$ no longer has to be diagonal in the 1,2 block.
It is this non-diagonal nature that allows a Cabibbo mixing of the first two generations to take place. 
Neither $F_b$ nor $F_a$ has that property so they cannot be used as residual
symmetry operators unless we abandon Cabibbo mixing.

Now that we know the residual symmetry to be $F_c$ in the $e$ sector and $G_3$ in the $\nu$ sector, we can use \eq{c2}, \eq{fgvev}, and \eq{IR} to compute the 
vacuum alignments. The result is
\be\begin{array}{|c|ccccc|} \hline &A&B&C&D&E\\ \hline
\bk{\Phi}&1&0&1&0&(0,1)\\
\bk{\Psi}&1&1&0&0&(1,-1)\\
\hline\end{array}\label{tvev}\ee
Since we know the left-handed fermions to belong to $D\oplus E$, \eq{IR} can be used to determine what irreducible multiplets of $\Phi$ and $\Psi$ 
the fermions can be coupled to 
in $\L_{eff}$. Taking into account \eq{tvev} which compels some allowed couplings to effectively vanish after symmetry breaking,
we end up with the following coupling schemes of  $e, u, d$ to $\bk{\Phi}$ (left-hand table) and $\nu$ to $\bk{\Psi}$ (right-hand table),
\be\begin{array}{|c|cc|c|c|cc|}\cline{1-3} \cline{5-7} \bk{\Phi}&D&E&\hspace{2cm}&\bk{\Psi}&D&E\\ \cline{1-3} \cline{5-7}
D&a'&e(1,0)&&D&\alpha'&\epsilon(1,-1)\\
E&e(1,0)&a,0,c,0&&E&\epsilon(1,-1)&\alpha,\beta,0,0\\
\cline{1-3} \cline{5-7}\end{array}\quad,\quad \label{elec}\ee
where the lower-case Latin and Greek letters are the coupling constants to the Higgs field of an irreducible representation indicated by the corresponding
upper case letter. For example, 
$a, a', \alpha$ and $\alpha'$ are coupling constants for irreducible representation $A$. More specifically, 
$\alpha'$ is the coupling constant of the $DD{\bf A}$ term, and $\alpha$ is the coupling constant 
of the $EE{\bf A}$ term, where the bold 
letter represents the Higgs representation, and the other two letters represents the fermions (remember the first generation belongs to $D$,
the second and third generations belong to $E$). $\gamma$ and $\delta$ never appear because $\bk{\Psi}=0$ for ${\bf C}$ and ${\bf D}$. Similarly, $b$ and
$d$ never occurs  because $\bk{\Phi}=0$ for ${\bf B}$ and ${\bf C}$, according to \eq{tvev}.

We have sufficient information now to write down $\L_{eff}$ and then the mass matrices under this $D_4$ scheme. The result is 
\be \M_i=\bp a_i'&e_i&0\\ e_i^*&a_i+c_i&0\\ 0&0&a_i-c_i\\ \ep\ (i=e,u,d),\qquad \M_\nu=\bp \alpha'&\epsilon&\epsilon\\ \epsilon&\alpha&\beta\\ \epsilon&\beta&\alpha\ep.
\label{mass}\ee
If we set the parameter $e_e=0$, then $\M_e$ is diagonal, and $a'_e, a_e, c_e$ can be used to fit the charged-lepton masses. $\M_\nu$ is 2-3 symmetric, has an 
invariant eigenvector $(0, 1, -1)^T$, hence neutrino mixing is automatically bimaximal. This is expected because $G_3$ is a residual symmetry. The remaining four
parameters in $\M_\nu$ can be used to fit the neutrino masses and the remaining mixing angles. For example, if we set $\alpha'+\epsilon=\alpha-\beta$, then
$(1,1,1)^T$ is an invariant eigenvector so trimaximal mixing is present. In that case the mixing matrix is tri-bimaximal, and the remaining three
parameters can be used to fit the neutrino masses. In the quark sector, if we set $e_d=0$,
then $\M_d$ is diagonal, and $a'_d, a_d, c_d$ can be used to fit the down-quark masses. $e_u$ can then be used to fit the Cabibbo angle, and $a'_u, a_u, c_u$
the up-quark masses. We see therefore that $\G=D_4$ can accommodate both tri-bimaximal neutrino mixing and the Cabibbo mixing of quarks with the same set of
Higgs, without giving up the criteria \eq{c1} and \eq{c2}.

This conclusion is reached without having to know how the right-handed fermions are doing. If we want a model of the local Lagrangian $\L$, all we have to do
is to assign the right-handed fermions to have the same representation as the left-handed fermions, namely, $D\oplus E$, then everything else is the same,
and the resulting mass matrices $M_\alpha$ will have the same form as $\M_\alpha$ in \eq{mass}.


\end{document}